\documentclass[floatfix,tightenlines,twocolumn,amsmath,amssymb,footinbib]{revtex4-1}

\usepackage[caption=false]{subfig}
\usepackage{graphicx}
\usepackage{bbm}
\usepackage{tensor}

\newcommand{\bra}[1]{\ensuremath{\left\langle #1\right\vert}}
\newcommand{\ket}[1]{\ensuremath{\left\vert #1\right\rangle}}
\newcommand{\expval}[1]{\ensuremath{\left\langle #1 \right\rangle}}
\newcommand{\hsp}[1]{\hspace{#1 em}}
\newcommand{\sqz}{\hsp{-0.1}}
\newcommand{\ketbra}[2]{\left\vert{#1}\right\rangle \sqz\sqz\sqz \left\langle{#2}\right\vert}

\newcommand{\tr}[1]{\mathrm{Tr}(#1)}

\def\iden{\mathbbm{1}}

\newcommand{\ezero}{\ket{\varepsilon_0}}
\newcommand{\eone}{\ket{\varepsilon_1}}
\newcommand{\etwo}{\ket{\varepsilon_2}}
\newcommand{\ethree}{\ket{\varepsilon_3}}
\newcommand{\eonen}{\ket{\varepsilon_1;n}}
\newcommand{\etwonp}{\ket{\varepsilon_2;n+1}}
\newcommand{\etwon}{\ket{\varepsilon_2;n}}
\newcommand{\invcm}{~\textrm{cm}^{-1}}

\graphicspath{{figures/}}

\begin{document}
\title{Enhancing light-harvesting power with coherent vibrational interactions:\\ a quantum heat engine picture}
\date{\today}

\author{N. Killoran, S.~F. Huelga, and M.~B. Plenio}
\affiliation{Institut f\"{u}r Theoretische Physik, Albert-Einstein-Allee 11, Universit\"{a}t Ulm, D-89069 Ulm, Germany}

\begin{abstract}
Recent evidence suggests that quantum effects may have functional importance in biological light-harvesting systems. Along with delocalized electronic excitations, it is now suspected that quantum coherent interactions with certain near-resonant vibrations may contribute to light-harvesting performance. However, the actual quantum advantage offered by such coherent vibrational interactions has not yet been established. We investigate a quantum design principle, whereby coherent exchange of single energy quanta between electronic and vibrational degrees of freedom can enhance a light-harvesting system's power above what is possible by thermal mechanisms alone. We present a prototype quantum heat engine which cleanly illustrates this quantum design principle, and quantify its quantum advantage using thermodynamic measures of performance. We also demonstrate the principle's relevance in parameter regimes connected to natural light-harvesting structures.
\end{abstract}

\maketitle


\section{Introduction}

The primary insight of quantum information science is that, by exploiting quantum effects, we can enhance performance over comparable classical devices. So it is humbling to think that nature may have taken advantage of quantum effects long before we had the idea. Indeed, there are several scenarios where quantum phenomena could play a role in biological function. One instance is biological light-harvesting; growing experimental evidence \cite{brixner05a,engel07a,calhoun09a,collini10a,panitchayangkoon10a,harel12a,romero14a,fuller14a} and theoretical models \cite{plenio08a,mohseni08a,caruso09a,ishizaki09a,chin10a,delrey12a, fuller14a} now suggest that quantum processes can contribute beneficially to the high efficiency of biological light-harvesting complexes (LHCs) (see \cite{huelga13a} for a recent review). Identifying potential quantum-mechanical design principles in such natural systems can, in turn, inspire new artificial light-harvesting technologies \cite{scholes11a,scully10a,scully11a,chin13b,zhang14a,
ajisaka14a}.

Attention has focused of late on the role of latent vibrational structures in LHCs \cite{womick11a, caycedo12a, chin12a, christensson12a, chin13a, plenio13a, huelga13a, kolli12a}. While most vibrations can be coarse-grained into a background environment, it is now thought that a discrete subset of strongly coupled modes should be considered on the same footing as the electronic light-harvesting system itself. Models where electronic excitations interact strongly with near-resonant underdamped vibrations \cite{chin12a,christensson12a,chin13a,tiwari13a,chenu13a,plenio13a} can explain the unusually long-lived coherence observed in photosynthetic complexes \cite{engel07a,calhoun09a,collini10a,panitchayangkoon10a,harel12a} and can exhibit enhanced transport properties \cite{womick11a,kolli12a,hossein12a,mourokh14a,irish14a,lim14a,oreilly14a}. Interestingly, with such a model, the vibrations can develop unambiguous signatures of non-classicality \cite{oreilly14a}, which, in examples, correlate suggestively with energy
transport. But despite these intriguing observations, the `quantum advantage' to light-harvesting provided by non-classical vibrations has not yet been fully elucidated.

In this work, we present a quantum design principle in action, whereby coherent interactions between electronic excitations and surrounding near-resonant vibrations lead to enhanced light-harvesting capabilities for the collective system. Taking a thermodynamic perspective, we focus on quantumness in the light-harvesting dynamics rather than in the states. Combining typical network transport models with a quantum heat engine picture \cite{scovil59a,shockley61a,alicki79a}, we present a prototype light-harvester which clearly reveals the vibrational quantum advantage. Specifically, near-resonant quantized vibrations have a catalytic role, opening up new energy transfer pathways and allowing the light-harvesting process to occur at a higher rate. These additional pathways are intrinsically quantum mechanical, involving coherent energy exchange between the electronic and vibrational subsystems via a Jaynes-Cummings interaction. Even with limited coherence, these processes can \emph{outrace} thermal
mechanisms at natural conditions, leading to larger energy currents and an overall enhancement in light-harvesting power. 

After presenting the basic design principle with the aid of an idealized prototype model, we examine the question of whether the observed behaviour is still present in non-ideal situations. Indeed, in realistic scenarios, both natural and artificial, there will be many deviations and imperfections which could wash out any performance improvements. Specifically, we analyse the effects of having off-resonant vibrations, non-optimal electronic couplings, and additional sources of decoherence. While any of these could be deleterious to performance, we show that the vibronic-induced enhancements remain even in such non-ideal settings. Finally, in addition to our idealized light-harvester, we will also demonstrate these principles in action within a parameter regime inspired by biological light-harvesting complexes, specifically the Cryptophyte algae PC645 complex.


\section{Light-harvesting framework}

\subsection{Quantum heat engine model}\label{ssec:qhem}

\begin{figure}[!t]
    \subfloat[]{%
	\includegraphics[width=0.45\columnwidth]{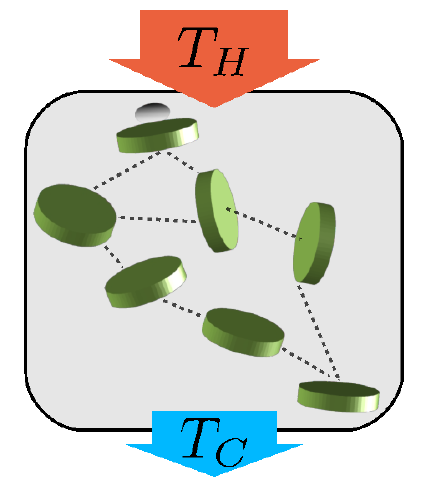}
    }
    \hfill
    \subfloat[]{%
	\includegraphics[width=0.45\columnwidth]{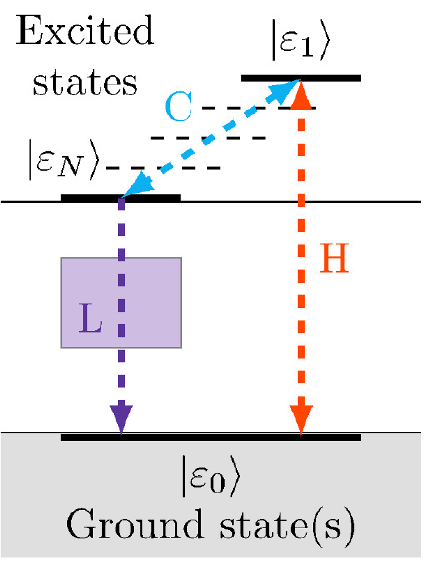}
    }
    \caption{Basic framework for light-harvesting: a) Incoming energy (upper arrow) produces an excited electron in a transport network. The combined effect of coherent (dotted lines) and dissipative processes (lower arrow) transfer the excitation through the network. b) The electronic eigenstates partition into two subsets separated by an energy gap. Considering an abstract load between the upper and lower `terminals,' we can study the system's power-delivery capabilities.}
    \label{fig:basicscheme}
\end{figure}

Both natural and artificial light-harvesting can be understood within a common framework (Fig. \ref{fig:basicscheme}a). Two subsets of electronic states (ground/excited subspaces in LHCs; valence/conduction bands in semiconductors) are separated by a large energy gap $E_g$. Incoming energy excites an electron out of a low-energy state, across the energy gap, into a high-energy excited state. In semiconductor solar cells, this energy comes directly from sunlight, and many different excited states could be populated. In contrast, for organic systems, the photosynthetic apparatus can have many distinct sub-units, with many segments receiving energy thermally from their neighbours, not directly from the sun. In the following, we take a specific state, e.g., the one with highest energy, to be the only one which gets initially excited (this starting point can be justified along the lines of \cite{chin13b}, where two coupled dipoles can lead to the formation of an active ``bright'' state and an inactive ``dark'' state).

The excitation is transported through a network of excited states to the edge of the gap. Here, we picture an abstract black-box load bridging the separated electronic `terminals.' Similar to an electric circuit, the load could be its own functional unit, such as an LHC's reaction center or an electrochemical battery, or it could represent some further subnetwork of transport paths. By transferring its excess energy to the load system, the electronic system can return to the low-energy subspace. Note that in semiconductor systems, the electron is physically transported, while in biological systems, it is an electronic excitation (exciton) that is transported, with electrons remaining on localized sites.

We consider the single excitation subspace, with corresponding electronic Hamiltonian
\begin{equation}\label{eq:He}
H_e= E_0\ketbra{\varepsilon_0}{\varepsilon_0} + \sum_{i=1}^N E_i\ketbra{s_i}{s_i} + \sum_{i\neq j}J_{ij}\ketbra{s_i}{s_j}.
\end{equation}
The orthogonal states $\{\ket{s_i}\}$ correspond to excitations spatially localized at specific sites within the network. For our considerations, the low-energy subspace will contain only the ground state $\ket{\varepsilon_0}$. The distinction into separate spatially localized sites is motivated by physical geometries. However, in many situations of interest, the couplings $J_{ij}$ between different sites are non-zero, leading to delocalized exciton eigenstates $\{\ket{\varepsilon_i}\}_{i=0}^N$, with energies $\varepsilon_i$. Such delocalization can be remarkably robust, even in noisy conditions \cite{caruso09a,caruso10a,fassioli14a}.

Our light-harvester's evolution will be described by a Lindblad-type master equation \cite{lindblad76a,gorini76a,rivas11a},
\begin{equation}
	\dot{\rho}_S(t) = i[\rho_S(t),H_S]+\sum_\alpha\mathcal{L}_\alpha[\rho_S(t)].\label{eq:mastereqn}
\end{equation}
Each Lindblad superoperator $\mathcal{L}_\alpha$ is associated with a transition rate $\Gamma_\alpha > 0$ and a transition operator $A_\alpha$:
\begin{equation}
	\mathcal{L}_\alpha(\rho_S) = \Gamma_{\alpha} [A_\alpha\rho_S A_\alpha^\dagger - \tfrac{1}{2}\{A_\alpha^\dagger A_\alpha,\rho_S\}].
\end{equation}
While other types of equations exist for describing open system evolution (e.g., Redfield equations), Lindblad-type master equations have the desirable feature that they guarantee completely positive (i.e., physically consistent) evolution of the quantum system \cite{lindblad76a,gorini76a,rivas11a}. For one, this means that no population can ever become negative, an issue which has been a catch in previous heat engine models based on perturbative equations (as pointed out in \cite{chin13b}). Another advantage of Lindblad evolution versus perturbative approaches is that it also forces coherences (i.e., off-diagonal elements of the density matrix) to evolve in a physically consistent way.

We analyze our light-harvester as a \emph{quantum heat engine} (QHE) \cite{shockley61a,scully03a,scully10a,scully11a,dorfman13a,chin13b,oreillyThesis}. This perspective has revealed functional roles for quantum coherence in both natural and artificial systems. The present work can be seen as an extension of these ideas to the case of mixed electronic-vibrational (i.e., vibronic) coherence. QHE performance is determined by the rate at which useful energy is transferred to the load, i.e., by the power $P=I_LV_L$, where $I_L$ and $V_L$ are the current and voltage across the load. The current is simply $I_L=e\rho_{NN}\Gamma_L$, with $\rho_{NN}$ the population of the lowest excited state ($e$ is the electron charge). The voltage is given by $eV_L=E_g+k_B T_C \ln(\rho_{NN}/\rho_{00})$, where $E_g=\varepsilon_N-\varepsilon_0$ and $k_B$ is Boltzmann's constant. This quantifies the energy that could be released by equilibrating a given population ratio at temperature $T_C$. Using detailed balance arguments, Shockley and Queisser \cite{shockley61a}
 showed that the maximum achievable voltage of such gapped systems is $eV_L\leq \eta_C E_g$,
where $\eta_C:=1-{T_C}/{T_H}$ is the Carnot efficiency. Since $V_L$ is fundamentally limited, the best strategy to increase power may therefore be to increase $I_L$, which is proportional to the population $\rho_{NN}$. The more population in this lowest excited state during operation, the more power the heat engine can deliver. As we will soon see, this is the key advantage offered by coherently interacting electronic and vibrational systems.

The heat engine viewpoint lets us consider the detailed inner workings of the light-harvester, while simplifying the sources, sinks, and load to which it connects. The QHE `system' degrees of freedom are included in the density matrix $\rho_S$ and the (time-independent) Hamiltonian $H_S$. 
The QHE connects to three important external systems: a `hot' source of incoming energy, a `cold' dissipative background, and a black-box load (see Fig. \ref{fig:basicscheme}b). We picture each of these external systems as thermal reservoirs at temperatures $T_H$, $T_C$, and $T_L$, respectively, with $T_H>T_C$. For simplicity, we assume the load transition is unidirectional, and model it using $T_L=0$. This means that only the heat engine can transfer energy to the load, not the other way around. To best visualize the dynamics, we will consider the electronic level scheme in the exciton basis $\{\ket{\varepsilon_i}\}_{i=0}^N$; thus, the electronic Hamiltonian $H_e$ alone cannot cause transitions between these levels. Aside from the ground state $\ket{\varepsilon_0}$, the electronic eigenstates are ordered by their energy ($\eone$ highest, $\ket{\varepsilon_{N}}$ lowest).

The three reservoirs facilitate all possible transitions between the exciton states $\ket{\varepsilon_i}$. The hot reservoir drives transitions only between the ground state $\ket{\varepsilon_0}$ and the highest energy exciton state $\eone$. Conversely, the load transfers the exciton only from the lowest excited state $\ket{\varepsilon_N}$ back to the ground state $\ket{\varepsilon_0}$. On the other hand, the dissipative cold background reservoir causes transitions between all pairs of excitons. This includes not only beneficial transitions $\ket{\varepsilon_i}\rightarrow\ket{\varepsilon_j}$ ($i,j\neq 0$) which assist transport through the excited state network, but also deleterious transitions $\ket{\varepsilon_i}\rightarrow\ket{\varepsilon_0}$ ($i\neq 0$) which represent unwanted but physically important decays of excitons back to the ground state without giving their energy to the load. It is exactly these loss mechanisms which a light-harvesting system is competing against. The lossy transitions will occur at a fixed rate, so the faster excitation energy can be safely transported through the network, the better. The interaction of the QHE with each of the thermal reservoirs is parameterized by two quantities, namely a coupling 
$\gamma_\alpha$, $\alpha\in\{H,C,L\}$, and the reservoir's mean occupation number $n_{\alpha ij}$, $i,j\in\{0,1,\dots,N\}$, at the given transition energy $\varepsilon_{ij}=\varepsilon_i-\varepsilon_j$. Because the hot reservoir and the load only link two levels, we can omit the redundant $ij$ indices for these. The full set of employed Lindblad terms is found in Table \ref{tab:lindblads}. 
\begin{center}
	\begin{table}
	\begin{tabular}{|c|c|c|}
		\hline
		& $\{\mathrm{transition}~A_\alpha,~\mathrm{rate}~\Gamma_\alpha\}$ & $\{\mathrm{transition}~A_\alpha,~\mathrm{rate}~\Gamma_\alpha\}$ \\
		& (forward transition) & (reverse transition)\\
		\hline
		$\mathcal{L}_{H}$: & $\{\ketbra{\varepsilon_1}{\varepsilon_0},\gamma_H\overline{n}_H\}$, & $\{\ketbra{\varepsilon_0}{\varepsilon_1},\gamma_H(\overline{n}_H+1)\}$\\
		$\mathcal{L}_{Cij}$: &  $\{\ketbra{\varepsilon_j}{\varepsilon_i},\gamma_{C}(\overline{n}_{Cij}+1)\}$, & $\{\ketbra{\varepsilon_i}{\varepsilon_j},\gamma_{C}\overline{n}_{Cij}\}$\\		
		$\mathcal{L}_{L}$: & $\{\ketbra{\varepsilon_0}{\varepsilon_N},\Gamma_{L}\}$ & \\
		\hline
	\end{tabular}
	\caption{List of transition operators $A_\alpha$ and associated rates $\Gamma_\alpha$ used in our prototype model. The ``cold'' reservoir transitions $\alpha={Cij}$ occur for all distinct pairs $i,j$ of excitonic states.}\label{tab:lindblads}
	\end{table}
\end{center}
 
Except for $\mathcal{L}_H$, all forward transitions proceed to lower energies, indicating the desired directionality for the QHE circuit. For the hot reservoir, we fix the occupation $\overline{n}_H=60000$, to match with previous work \cite{dorfman13a,chin13b}. Importantly, we remark that this represents solar energy which has been concentrated within the larger antenna complex, and so the associated temperature is not an actual physical temperature. The thermal occupations of the cold reservoir are determined from the Planck distribution
\begin{equation}\label{eq:planckdist}
	\overline{n}_{Cij} = \frac{1}{\exp(\varepsilon_{ij}/(k_B T_C))-1}.
\end{equation}
Finally, different impedances of the load are modelled by varying the parameter $\Gamma_L$.


\subsection{Including the vibrational subsystem}

In addition to the electronic degrees of freedom, vibrations are important in light-harvesting systems. A full microscopic model would allow for many independent vibrational modes at each site, which would give an overall mode Hamiltonian
\begin{align}\label{eq:full_Hm}
H_m^\mathrm{full} = \sum_{i=1}^N\sum_\kappa \hbar\omega_{i\kappa} \hat{a}^\dagger_{i\kappa}\hat{a}_{i\kappa}.
\end{align}
These modes couple to the electronic system via a linear interaction
\begin{align}\label{eq:full_HI}
H_I^\mathrm{full} = \sum_{i=1}^N\sum_\kappa \hbar g_{i\kappa}\ketbra{s_i}{s_i}\otimes(\hat{a}_{i\kappa}+\hat{a}_{i\kappa}^\dagger),
\end{align}
where $g_{i\kappa}$ is the coupling constant for each mode. Motivated by recent work \cite{chin12a,caycedo12a,christensson12a,chin13a,tiwari13a,chenu13a,plenio13a,womick11a,kolli12a,hossein12a,mourokh14a,irish14a,lim14a,oreilly14a}, we consider a small collection of strongly coupled vibrations as part of the functional system. Fitting with our heat engine picture, we suppose that all remaining modes are part of the background thermal reservoir. We assume that the strongly-coupled modes are the only significant features in the environment. Since these are accounted for as part of the functional system, the remaining background will be featureless, and we will model it as a simple bosonic bath of temperature $T_C$ and uniform coupling $\gamma_C$ to the system.

Our basic prototype will have $N=3$ sites. To best demonstrate the design principle, we keep a single quantized mode each at sites 1 and 2 (with ladder operators $\hat{a}_{1/2}$ and frequencies $\omega_{1/2}$) as part of our functional system. We assume that there is no direct interaction between vibrations on different sites. Thus, the system Hamiltonian $H_S$ consists of three parts: the electronic Hamiltonian
\begin{equation}\label{eq:He_pt}
H_e= E_0\ketbra{\varepsilon_0}{\varepsilon_0} + \sum_{i=1}^2 E_i\ketbra{s_i}{s_i} + \sum_{i\neq j}J_{ij}\ketbra{s_i}{s_j},
\end{equation}
a single mode Hamiltonian at sites 1 and 2,
\begin{align}\label{eq:Hm_pt}
H_m = \sum_{i=1}^2\hbar\omega_{i} \hat{a}^\dagger_{i}\hat{a}_{i},
\end{align}
and a single interaction term at each site, 
\begin{align}\label{eq:HI_pt}
H_I = \sum_{i=1}^2\hbar g_{i}\ketbra{s_i}{s_i}\otimes(\hat{a}_{i}+\hat{a}_{i}^\dagger).
\end{align}
Assuming similar structure at each site, we fix $\omega_1=\omega_2 = \omega_m$ and $g_1=g_2=g$ to have the same values for each mode. The full system Hamiltonian $H_S := H_e+H_m+ H_I$ describes the unitary part of the evolution in Eq. (\ref{eq:mastereqn}). We note that the quantum design principle also works in less idealized scenarios (see Sec. \ref{sec:nonideal}), as well as larger networks.

To include the quantized modes numerically, we truncate the Hilbert space of each mode at some fixed dimension $D=N_\mathrm{max}+1$. For all our examples, we use $D=6$. Physically, the modes should also undergo relaxation, though at a somewhat slower rate than the electronic subsystem. Accordingly, we include additional Lindblad terms $\mathcal{L}_{Mi}^{\downarrow/\uparrow}$ for each quantized mode. As with the excitons, these terms represent local interaction with a thermal reservoir, and are parameterized by couplings $\gamma_M$ and mean bath occupations $\overline{n}_M$ (assumed to be the same at each site). The Lindblad rates are thus $\Gamma_{Mi}^\downarrow=\gamma_M(\overline{n}_M+1)$ (damping) and $\Gamma_{Mi}^\uparrow=\gamma_M\overline{n}_M$ (excitation), while the transition operators are simply the ladder operators $A_{Mi}^\downarrow = \hat{a}_i$ (damping)
and
$A_{Mi}^\uparrow = \hat{a}_i^\dagger$ (excitation). We have set $\gamma_M$ to give a rate $1/(1~\textrm{ps})$. The occupation numbers $\overline{n}_M$ are calculated using a Planck distribution (Eq. (\ref{eq:planckdist})) at temperature $T_M$ for a level spacing $\varepsilon_{i+1,i}=\hbar\omega_m$.
The specific numerical parameter values for all examples are listed in Table \ref{tab:modelparameters}.

\begin{center}
\begin{table}
	\begin{tabular}{| c | c | c | c |}
		\hline
		  & Prototype & PC645\\
		\hline
		$E_1$ & $300$ & $1226$ \\
		$E_2$ & $300$ & $1145$  \\
		$E_3$ & $0$ & $0$ \\
		$E_0$ & $-10000$ & $-15888$ \\
		$J_{12}=J_{21}$ & $100$ & $319.4$ \\
		$J_{13}=J_{31}$ & $0$ & $0$ \\
		$J_{23}=J_{32}$ & $0$ & $0$ \\
		$\hbar\omega_m$ & $200$ & $807$ \\
		$\hbar g$ & $55$ & $200$ \\
		\hline
		$\hbar\gamma_H$ & $0.01$ & $0.01$ \\
		$\hbar\gamma_C$ & $8.07$ & $24.4$ \\
		$\hbar\gamma_M$ & $5.3$ & $5.3$ \\
		$\overline{n}_H$ & $60000$ & $60000$ \\
		$T_C$ &  $293$ & $293$ \\
		$T_M$ & $293$ & $293$ \\
		\hline
	\end{tabular}\caption{Numerical parameters used in this paper. Energies, couplings, and rates are in units of $\invcm$, mean phonon numbers are unitless, and temperatures are in K. For analytical simplicity, we have taken the couplings $J_{13}$ and $J_{23}$, assumed small compared to the other Hamiltonian parameters, to be zero. In a detailed microscopic model, these couplings would be small but finite. We capture these finite couplings instead via our Lindblad transition model.}
	\label{tab:modelparameters}
\end{table}
\end{center}

While the Hamiltonian eigenstates are modified by vibrational coupling, we assume that the non-unitary part of the master equation is not significantly changed by vibronic mixing. Thus, we use the same Lindblad terms both with and without electron-mode coupling. A similar approach is commonly encountered in other physical systems using the Jaynes-Cummings interaction, such as cavity quantum electrodynamics. There, non-mixed Lindblad terms can give accurate predictions of the dynamics provided two conditions are met \cite{scala07a, scala07b}, namely that the Hamiltonian transition frequencies are much larger than the dissipative decay rates, and that the environmental spectral density is relatively flat and featureless. Such conditions are present in our model (see, e.g., Table \ref{tab:modelparameters}), so the adopted phenomenological framework is well motivated.

\section{Vibronic light-harvester prototype}

Motivated by values found in biological systems, we take site energies $E_1 = E_2 = 300 \invcm$, $E_3 = 0 \invcm$, and ground state energy $E_0=-E_g = -10000\invcm$. The two highest excited states are coupled, $J_{12}=J_{21} = 100\invcm $, while the remaining couplings $J_{13}=J_{31}$ and $J_{23}=J_{32}$ are small in comparison.
A suitable basis for analyzing energy transfer is thus the eigenbasis with $J_{13}=J_{23}=0$ 
(the small but finite character of the couplings is incorporated phenomenologically within the Lindblad transitions). These values lead to electronic eigenstates $\{\ket{\varepsilon_i}\}_{i=0}^3$, where $\ket{\varepsilon_{1/2}}=\frac{1}{\sqrt{2}}[\ket{s_1}\pm \ket{s_2}]$ are delocalized, while $\ethree=\ket{s_3}$ and $\ezero$ is unchanged. The corresponding energies are $\varepsilon_{1/2}=E_{1/2}\pm J_{12} = 300 \pm 100 \invcm$, with $\varepsilon_0=E_0$ and $\varepsilon_3=E_3$ as before. Finally, we set the mode frequency
resonant with delocalized exciton spacing $\hbar\omega_m = \varepsilon_1-\varepsilon_2 = 200 \invcm$.

\subsection{Analysis of energy transfer dynamics}

In this subsection, we explore the various energy transfer processes of the prototype model. A good grasp of these competing mechanisms can help us understand the behaviour of the system at the steady state. The Lindblad transitions were already discussed in section \ref{ssec:qhem}, so we focus here on the new unitary contribution arising from the joint electron-mode Hamiltonian. In the given parameter regime, the interaction becomes $H_I = H_{CM} + H_{RD}$, with
\begin{align}
	H_{CM} = & ~\frac{\hbar g}{\sqrt{2}}(\ketbra{\varepsilon_1}{\varepsilon_1}+\ketbra{\varepsilon_2}{\varepsilon_2})\otimes(\hat{a}_{CM}+\hat{a}_{CM}^\dagger),\\
	H_{RD} = & ~\frac{\hbar g}{\sqrt{2}}(\ketbra{\varepsilon_1}{\varepsilon_2}+\ketbra{\varepsilon_2}{\varepsilon_1})\otimes(\hat{a}_{RD}+\hat{a}_{RD}^\dagger).
\end{align}
In these mutually commuting terms, we have introduced centre of mass/relative displacement modes $\hat{a}_{CM/RD}=\frac{1}{\sqrt{2}}(\hat{a}_1\pm\hat{a}_2)$. Notice that both vibronic coupling \emph{and} delocalized excitons were important for arriving at this form. The term $H_{RD}$ is a Jaynes-Cummings (JC) interaction, whose eigenstates, within the rotating wave approximation (RWA), are $\{\ket{\varepsilon_2;0}, \ket{\pm;n}\}_{n=0}^{\infty}$, where the states $\ket{\pm;n}:=\tfrac{1}{\sqrt{2}}[\ket{\varepsilon_1;n}\pm\ket{\varepsilon_2;n+1}]$ are entangled between the electronic and vibrational subsystems (we use the shorthand  $\ket{\varepsilon_i;n}:=\ket{\varepsilon_i}\otimes\ket{n}_{RD}$). Jaynes-Cummings physics is decidedly non-classical in nature. In the absence of competing processes, a Jaynes-Cummings interaction can produce entanglement between electronic and mode subsystems, and can lead to mode states with non-classical properties (sub-Poissonian statistics, quadrature squeezing, and negative quasiprobabilities) \cite{shore93a}. 

\begin{figure}[!t]
\centering
\includegraphics[width=\columnwidth]{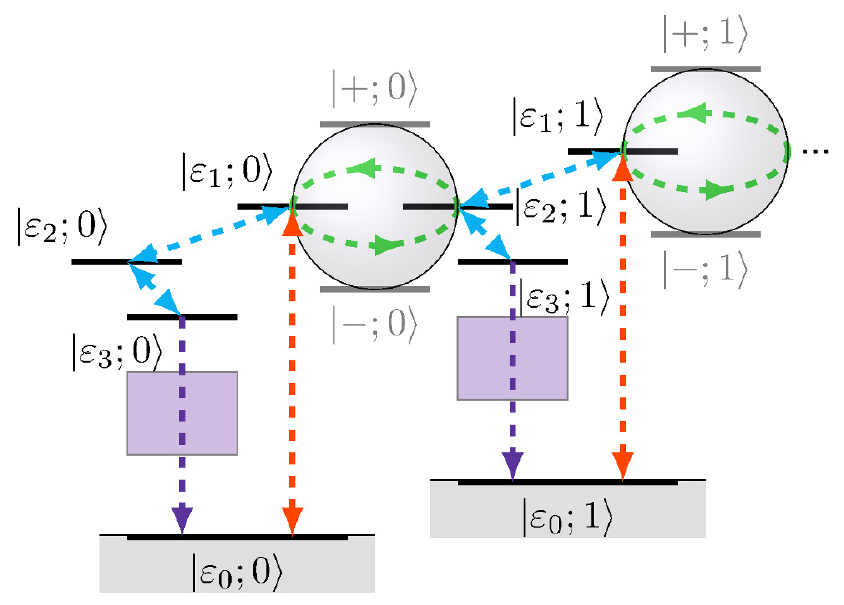}
\caption{Basic incoherent and coherent transfer mechanisms with respect to the uncoupled basis vectors $\{\ket{\varepsilon_i;n}\}$. The coupled Hamiltonian generates rotations around the axis between $\ket{+;n}$ and $\ket{-;n}$, depicted using Bloch spheres for each two-dimensional subspace. This manifests as coherent oscillations between the states $\ket{\varepsilon_1;n}$ and $\ket{\varepsilon_2;n+1}$. For clarity, some dissipative transitions, as well as damping of the mode, are not shown.}
\label{fig:energyflow1}
\end{figure}

The Jaynes-Cummings interaction induces coherent population transfer between the states $\eonen$ and $\etwonp$ at frequencies $\omega_\textrm{coh}(n)\sim 2\hbar g\sqrt{n+1}$. In isolation, this evolution would manifest as Rabi oscillations. This mechanism provides an additional channel for excitonic population to transfer between $\eone$ and $\etwo$. Consider the electron/mode state $\ket{\varepsilon_1;n}$. Without coherent interactions, the exciton would dissipate some of its energy to the background reservoir, relaxing to the nearby state $\etwo$ at the rate $\Gamma_{C12}$ (the transition $\eone\rightarrow\ethree$ is less likely because of the larger energy difference). On the other hand, the vibronic transfer $\eonen\rightarrow\etwonp$ provides a \emph{coherent shortcut} bypassing the thermal electron-only transition $\eonen\rightarrow\etwon$ (Fig. \ref{fig:energyflow1}), allowing $\etwo$ to be populated at earlier times. In turn, the load-connected state $\ethree$ will also be populated earlier.

\begin{figure}[!t]
\centering
\includegraphics[width=\columnwidth]{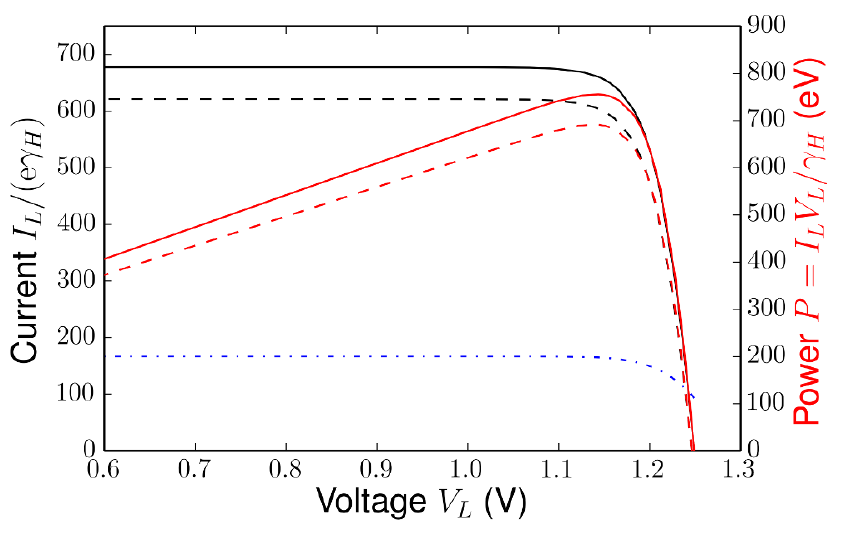}
\caption{Numerical I-V characteristic (upper black curve pair) and power (middle red curve pair) of our prototype quantum heat engine with (solid lines) and without (dashed lines) coherent vibrational interactions. The lower blue dash-dotted curve represents the magnitude of $I_\mathrm{coh}$ up to n=5. The maximum power is enhanced with vibrational coupling by $\sim 9.3\%$ relative to the uncoupled case.}
\label{fig:prototypeIV}
\end{figure}

Coherent exchange can also transfer population \emph{back} from $\ket{\varepsilon_2;n+1}\rightarrow\ket{\varepsilon_1;n}$. However, the thermal transition $\ket{\varepsilon_2;n+1}\rightarrow$ $\ket{\varepsilon_3;n+1}$ will transfer some population out of the two-level subspace, suppressing its revival in $\ket{\varepsilon_1;n}$ and enforcing directionality. Importantly, since the load is connected only to the electronic subsystem, it is insensitive to the mode. Therefore, energy can be transferred to the load even without the overall system completing a full thermodynamic cycle. By coherently absorbing single quanta of energy, the mode thus catalyzes faster exciton transfer through the network. Additionally, re-excitation of the electronic system can occur before the mode has dissipated the extra energy (this typically happens at a slower rate), allowing multiple mode levels to contribute in parallel. Over time, dissipative processes on the mode will regulate the mode population $\overline{n}_M$ (and thus the 
available oscillation frequencies $\omega_\mathrm{coh}(n)$), suppressing excessive back-transfer. At steady-state, the interplay of all these coherent and incoherent mechanisms
contributes to a higher overall load current $I_L$. In Fig. \ref{fig:prototypeIV} we plot the numerical steady-state I-V characteristic for our prototype, with and without vibrational coupling, demonstrating the enhanced power made possible by coherent vibronic evolution. 

We note that although the RWA is helpful for elucidating the underlying mechanisms of coherent energy transfer, all numerics are done using the full system Hamiltonian, without making the RWA. Even when the RWA is valid for a closed electron/vibration system, the evolution also contains a number of Lindblad transitions which will interact with the Hamiltonian evolution. The energy scales and interplay from the open system evolution may make the omitted RWA terms non-neglible, and we therefore use the unsimplified Hamiltonian in our simulations. 

We also emphasize that electron-vibration coupling alone does not lead to any population/energy transfer since the interaction Hamiltonian does not connect different sites. Excitonic energy transfer only becomes possible with this interaction if the electronic eigenstates have some amount of delocalization. Thus, delocalization can be thought of as a tool for `unlocking' transfer capabilities from the available interaction. In the prototype and the later cryptophyte example, only excitons 1 and 2 are delocalized, and a quantized mode at site $3$ would not contribute to the evolution. On the other hand, if the coupling between sites $2$ and $3$ was non-negligible, then having a quantized mode at site 3 could open up a second coherent pathway, which may further enhance transport.

Finally, to make the quantum design principle most clear, the prototype was designed to give an exactly resonant JC-like model where neither incoherent nor coherent processes dominate. In more realistic systems, things will not be so ideal, yet the principle still holds. Even resonance-detuned modes and partially delocalized excitons can combine to produce coherent oscillations. Whenever these oscillations have non-negligible overlap with $\eone$ and $\etwo$, they will provide beneficial alternate transfer pathways. In Sec. \ref{sec:nonideal} and the Appendix, we explore several more realistic models, including one where incoherent transport is already optimized, and the quantum design principle provides a further advantage. 


\subsection{Linking coherence to performance}

In the previous subsection, we explored the elementary mechanisms that the prototype light-harvester has available for energy transfer, and how these support or compete against one another. In this subsection, we will quantify how these processes, in particular coherent vibronic energy exchange via a Jaynes-Cummings-type interaction, determine the energy flows and thus performance at steady state. To quantify the influence of electron-vibration coherence, we define currents
\begin{align}
	I_{H} := & ~ \gamma_H[\rho^e_{00}(t)\bar{n}_H
	-\rho^e_{11}(t)(\bar{n}_H+1)],\label{eq:IH} \\
	I_{C}^{i\rightarrow j} := & ~ \gamma_C[\rho^e_{ii}(t)(\bar{n}_{Cij}+1)
	-\rho^e_{jj}(t)\bar{n}_{Cij}], \label{eq:ICij}\\
	I_{L} := & ~ \Gamma_L\rho^e_{33}(t).	
\end{align}
We track the flow of energy in our heat engine using the electronic Hamiltonian $H_e$. This flow contains two contributions, due to the unitary (Hamiltonian) and non-unitary (Lindbladian) parts of the master equation:
\begin{align}\label{eq:elecenergyflow}
	\tfrac{d}{dt}\expval{{H_e}} = i\tr{[\rho_S(t),H_S]H_e} +
	\tr{\mathcal{L}(\rho_S(t))H_e}.
\end{align}
The Lindbladian part takes the form
\begin{align}\label{eq:clheat2}
	\tr{\mathcal{L}(\rho_S)H_e}~
	= \underbrace{\varepsilon_{10}I_H}_{\dot{Q}_H}
	- \underbrace{\sum_{\{ij\}}\varepsilon_{ij}I_{C}^{i\rightarrow j}}_{\dot{Q}_{C}}
	- \underbrace{\varepsilon_{30}I_L}_{\dot{Q}_L},
\end{align}
where $\dot{Q}_\alpha$ are net energy flows and $\varepsilon_{ij}:= \varepsilon_i-\varepsilon_j$ are energy differences between the various states. Without coherent interactions, the excitonic populations $\rho^e_{kk}:=\tr{\rho_S\ketbra{\varepsilon_k}{\varepsilon_k}\otimes\iden_m}$ satisfy straightforward balance equations 
which dictate the value of the load current $I_L$ in steady state.

If the interaction $H_I$ is present, then $[H_S,H_e]\neq 0$, and there is an additional coherent energy exchange between the electronic system and the mode, $\dot{Q}_{e-m}:= i\tr{[\rho_S(t),H_S]H_e}$. Simplifying, we find
\begin{align}\label{eq:Qem}
	\dot{Q}_{e-m} = & ~ \varepsilon_{12}
	\left[
	\sqrt{2}g \sum_{n=0}^\infty\sqrt{n+1}~\textrm{Im}\Big(\rho_{{}\indices*{^2^{n+1}_1_n}}\Big)
	\right]
	=:\varepsilon_{12}I_\mathrm{coh},
\end{align}
with $\rho_{{}\indices*{^2^{n+1}_1_n}}:=\bra{\varepsilon_2;n+1}\rho_S(t)\ket{\varepsilon_1;n}$ and where we have defined the \emph{coherent current} $I_\textrm{coh}$ with the convention that $-I_\textrm{coh}>0$ when a net current flows from electron to mode. Clearly this is a non-classical flow of energy, arising from the Jaynes-Cummings interaction term in the Hamiltonian. In contrast to the Lindbladian terms, which are parameterized in terms of diagonal populations, this energy flow is mediated exclusively via off-diagonal coherences between the uncoupled basis vectors $\ket{\varepsilon_i;n}$. In fact, coherences from higher mode occupations $n$ contribute to the energy flow with more weight than lower ones, so even small amounts of electron/mode coherence can add significantly to energy transfer. 

The steady state energy currents can be determined by setting Eq. (\ref{eq:elecenergyflow}) to zero, giving
\begin{align}\label{eq:cohbalance}
	I_L = \frac{1}{\varepsilon_{30}}\left[
	\varepsilon_{10}I_H
	- \sum_{\varepsilon_i>\varepsilon_j}\varepsilon_{ij}I_{C}^{i\rightarrow j}
	+ \varepsilon_{12}(-I_\textrm{coh})
	\right].
\end{align}
Without vibrational interactions, the coherent current $I_\mathrm{coh}$ is zero and Eq. (\ref{eq:cohbalance}) becomes a standard balance relation for the currents. However, when $I_\mathrm{coh}\neq0$, the balance of currents can be modified. Vibronic coherence thus pushes the steady state away from the incoherent rate equation solution (involving only the terms $I_H$ and $I_C^{i\rightarrow j}$) expected by detailed balance arguments, thereby allowing the load current to be increased overall.


\section{Non-ideal situations}\label{sec:nonideal}

Our idealized prototype system is meant to reveal the design principle and its advantages most clearly. Yet it is important to show that the design principle can work even in less idealized situations. To this end, we will examine separately three sources of imperfection. These are quantized modes which are detuned from the excitonic transition frequencies, excitonic states which are only partially delocalized, and additional decoherence effects beyond those induced by the thermal transitions. Such imperfections could be caused, for example, by disorder (either static or dynamic) or by additional noise. We simplify the analysis by including the imperfections directly, rather than modelling their specific underlying causes.

For all imperfections, we take the same basic Hamiltonians as the prototype model (Eqs. (\ref{eq:He_pt})-(\ref{eq:HI_pt})). As before, the system evolution is described by a Lindblad master equation, with hot, cold, and load transitions occuring incoherently between the relevant excitonic eigenstates. Possible modifications of the system-bath interactions in these non-ideal cases, which should be based on a more detailed microscopic model, are beyond the scope of the present work. Finally, any parameter values which are not explicitly mentioned in the following examples are the same as in Table \ref{tab:modelparameters}.

\subsection{Off-resonant vibrations} One deviation from the ideal situation is when there are large mismatches between the exciton energy splitting and the frequency of the quantized vibrations, e.g., due to disorder. Such mismatches can suppress the coherent current, but will not destroy it altogether. To see this, we consider our prototype model, but with a mode frequency which is detuned by $\delta$ from the transition energy, i.e., $\hbar\omega_m=\varepsilon_{12}-\delta$. The interaction of Eq. (\ref{eq:HI_pt}) and the remaining parameters from Table \ref{tab:modelparameters}) are kept the same as the prototype.

\begin{figure}[!t]
\centering
\includegraphics[width=\columnwidth]{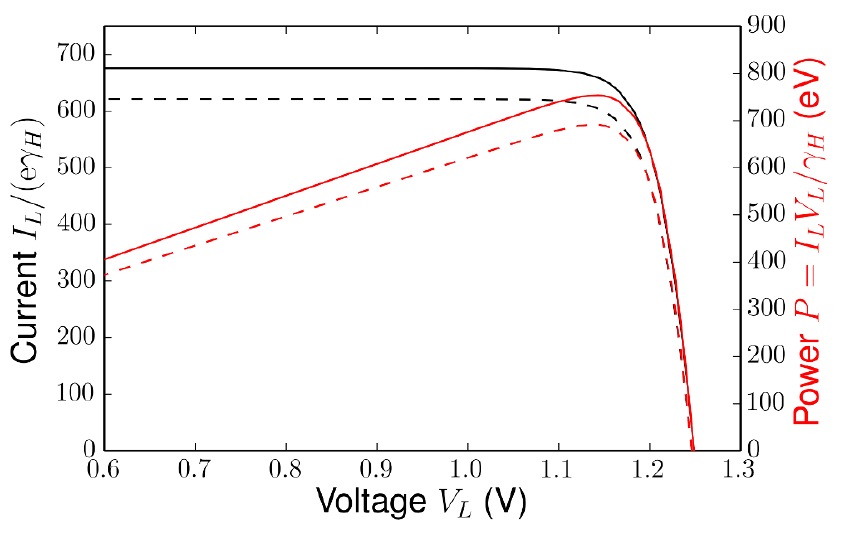}
\caption{Numerical I-V characteristic (upper black curves) and power (lower red curves) of a detuned model with (solid lines) and without (dashed lines) coherent exciton-mode interactions. Despite the detuning, the maximum power is still enhanced by $ \sim 9 \%$.}
\label{fig:detunedIV}
\end{figure}

When the rotating wave approximation is valid, the total Hamiltonian $H_S=H_e+H_m+H_I$ (a detuned Jaynes-Cummings model) can be diagonalized to obtain entangled vibronic eigenstates
\begin{align}
	\ket{+;n} = & \cos(\tfrac{\phi_n}{2})\eonen + \sin(\tfrac{\phi_n}{2})\etwonp\\
	\ket{-;n} = & \sin(\tfrac{\phi_n}{2})\eonen - \cos(\tfrac{\phi_n}{2})\etwonp
\end{align}
and eigenenergies
\begin{align}
	\varepsilon_{\pm n}:=\hbar\omega_m(n+\frac{1}{2})\pm \frac{1}{2}\Omega_n,
\end{align}
where $\Omega_n:=\sqrt{4\hbar^2 g^2(n+1)+\delta^2}$ is the Rabi frequency and $\tan(\phi_n)=2\hbar g\sqrt{n+1}/\delta$. Using the same intuition as the main text, for every level $n$, the Hamiltonian will cause coherent rotations on the Bloch sphere defined for the two states $\{\eonen,\etwonp\}$. In the ideal prototype, population was rotated directly along the equator; with detuning, the axis of rotation will point somewhere else on the sphere. As long as $\phi_n$ is not an integer multiple of $\pi$ (which happens when the coupling $g$ is zero or the detuning $\delta$ is infinitely far away), this rotation will still lead to partial population oscillations between the levels $\eonen$ and $\etwonp$. We also note that detuning modulates the rotation speed, which can in some situations lead to faster overall transfer even with a tilted axis of rotation. We plot the I-V and power curves of this detuned system in Fig. \ref{fig:detunedIV}, with
$\delta = 50 \invcm$, demonstrating that the design principle continues to yield a
power enhancement in this situation. As in the earlier section, we do not make the RWA in the actual numerics.

\subsection{Partial delocalization} 
\begin{figure}[!t]
\centering
\includegraphics[width=\columnwidth]{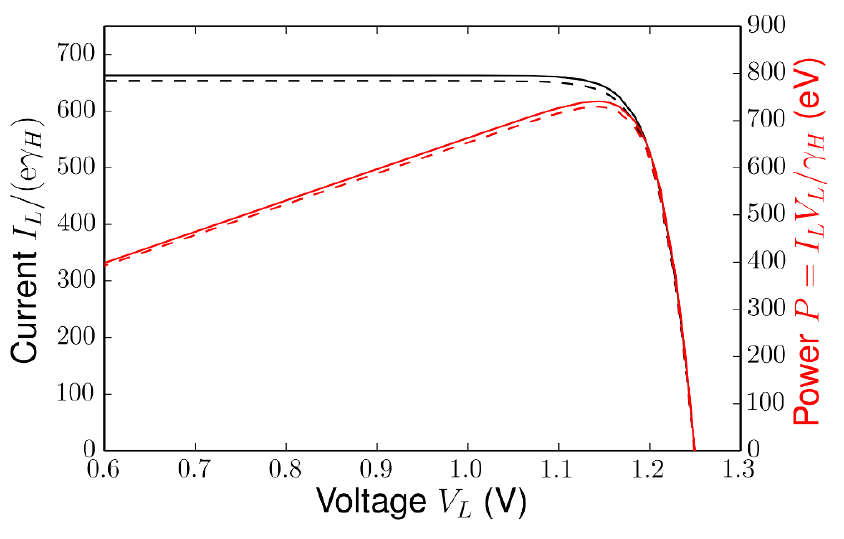}
\caption{Numerical I-V characteristic (upper black curves) and power (lower red curves) of a partially-delocalized prototype model with (solid lines) and without (dashed lines) coherent exciton-mode interactions. The design principle is still weakly at play in this non-ideal system, as evidenced by an enhancement in maximum power of $\sim 1.6\%$.}
\label{fig:partdeloc}
\end{figure}

Electronic delocalization was important in our prototype because it enabled the quantized vibrations to accept energy from the electronic subsystem under the given interaction. Delocalization effects can be examined by varying the dimer energies $E_{1/2}$ and coupling $J_{12}$ from the prototype model. In general, the eigenstates in the 1-2 subspace are
\begin{align}
\ket{\varepsilon_{1}} = & \cos(\tfrac{\theta_J}{2})\ket{s_1}+\sin(\tfrac{\theta_J}{2})\ket{s_2}\\ 	
\ket{\varepsilon_{2}} = & \sin(\tfrac{\theta_J}{2})\ket{s_1}-\cos(\tfrac{\theta_J}{2})\ket{s_2},
\end{align}
where $\tan(\theta_J) := \frac{2 J_{12}}{E_1-E_2}$ captures the degree of delocalization. The associated eigenenergies are $\varepsilon_{1/2}=\frac{1}{2}(E_1+E_2 \pm \sqrt{(E_1-E_2)^2+4J_{12}^2})$. With respect to these eigenstates, the interaction Hamiltonian becomes $H_I=H_{12} + H_{RD}$, where (using $\hat{X}_i := \hat{a}_i^\dagger + \hat{a}_i$)

\begin{align}
	H_{12} = & ~\hbar g\{
	\ketbra{\varepsilon_1}{\varepsilon_1}\otimes[
	\cos^2(\tfrac{\theta_J}{2})\hat{X}_1 
	+\sin^2(\tfrac{\theta_J}{2})\hat{X}_2]\ \nonumber\\
	& ~~ +\ketbra{\varepsilon_2}{\varepsilon_2}\otimes[
	\sin^2(\tfrac{\theta_J}{2})\hat{X}_1 
	+\cos^2(\tfrac{\theta_J}{2})\hat{X}_2]\} ,\\
	H_{RD} = & ~\frac{\hbar g}{\sqrt{2}}\sin(\theta_J)
	(\ketbra{\varepsilon_1}{\varepsilon_2}+\ketbra{\varepsilon_2}{\varepsilon_1})\otimes(\hat{a}_{RD}+\hat{a}_{RD}^\dagger).\label{eq:part_deloc_Ham}
\end{align}
So long as the delocalization angle $\theta_J$ is not an integer multiple of $\pi$ (which represents completely localized excitons), this interaction will still lead to coherent population oscillations between $\eone$ and $\etwo$ via Eq. (\ref{eq:mastereqn}), with the amplitude of the oscillations depending on the degree of delocalization through the prefactor $\frac{\hbar g}{\sqrt{2}}\sin(\theta_J)$. But we can see that a coherent current will be active, and the quantized modes contributing to the dynamics, for any non-zero amount of delocalization. In Fig. \ref{fig:partdeloc} we plot the I-V and power curves for our prototype with $E_1=310 \invcm$, $E_2=290 \invcm$, and $J_{12}=50 \invcm$. These values lead to partially delocalized excitons with $\cos(\tfrac{\theta_J}{2})\sim 0.83$ and $\sin(\tfrac{\theta_J}{2})\sim 0.56$, yet we still gain a power enhancement with the quantized modes. 

Both off-resonant vibrations and partial delocalization can be caused by disorder, either static or dynamic. Static disorder can lead to different energetic splittings and different degrees of excitonic delocalization, and hence different performance, within each realization. Dynamic disorder, on the other hand, will cause the configuration, and hence the enhancement, to change over time. However, unless the disorder completely relocalizes the excitons, the design principle will always be present to some degree. 

\subsection{Additional decoherence}

\begin{figure}[!t]
\centering
\includegraphics[width=\columnwidth]{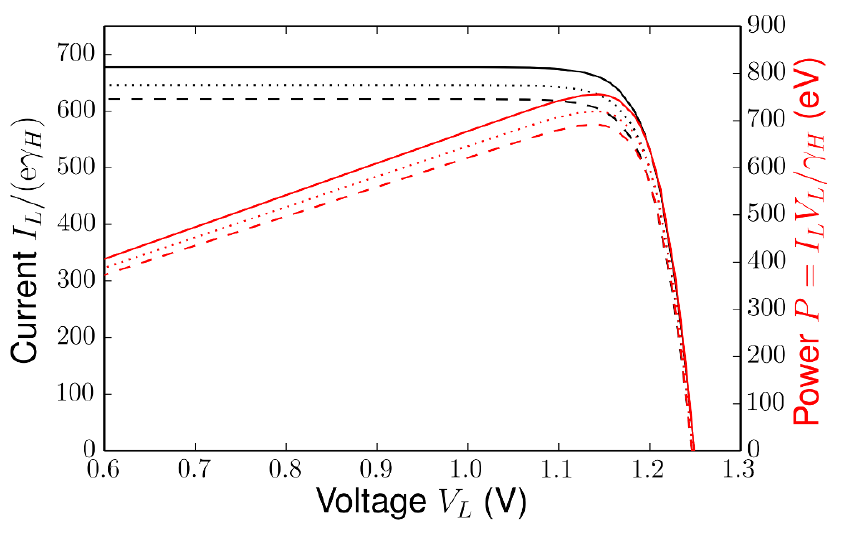}
\caption{Numerical I-V characteristic (upper black curves) and power (lower red curves) of a prototype model including extra decoherence. Solid/dashed lines are with/without coherent exciton-mode interactions, while the dotted line represents the case with both coherent interactions and decoherence.}
\label{fig:decoheredIV}
\end{figure}

Finally, we consider the influence of decoherence on our prototype. One potential source of decoherence has already been included in our model. Namely, strong interactions with the outside baths can suppress the flow of coherent current via quantum Zeno mechanisms, i.e., by applying the incoherent transition operators at too high a rate. Yet there may be other sources of decoherence present in the system, for instance due to microscopic details not accounted for in our master equation. To study this potential, we introduce an extra decoherence mechanism phenomenologically via a Lindblad term $\mathcal{L}_\mathrm{decoh}$ in the master equation. Since the design principle depends on coherence between the states $\eonen$ and $\etwonp$, the Lindblad operator is chosen to decohere in this basis. 

Specifically, we use
\begin{align}
	\mathcal{L}_\mathrm{decoh}= \gamma_\mathrm{decoh}\sum_{n=0}^{N_\mathrm{max}}\mathcal{L}_\mathrm{decoh}^{(n)},
\end{align}
where each $\mathcal{L}_\mathrm{decoh}^{(n)}$ has a transition operator
\begin{align}
A^{(n)} = \ketbra{\varepsilon_1;n}{\varepsilon_1;n} - \ketbra{\varepsilon_2;n+1}{\varepsilon_2;n+1}.	
\end{align}
Each of these operators is equivalent to a Pauli $\sigma_z$ in the corresponding basis $\{\eonen$ and $\etwonp\}$. Absent other dissipative terms in the evolution, they would cause pure dephasing, with the populations $\bra{\varepsilon_1;n}\rho\eonen$ and $\bra{\varepsilon_2;n+1}\rho\etwonp$ staying fixed while the off-diagonal elements $\bra{\varepsilon_1;n}\rho\etwonp$ decay exponential to zero. Since the coherent current of Eq. (\ref{eq:Qem}) is directly proportional to the imaginary part of these off-diagonal elements, these pure dephasing operators are directly opposing the new energy transfer pathway. We show the I-V and power curves for the decoherence rate $\gamma_\mathrm{decoh}=1.0$ eV in Fig. \ref{fig:decoheredIV}, where we can see that the power enhancement is only partially supressed by decoherence.


\section{Comparison of model} 

\subsection{Comparison to biological systems } 

\label{sec:PC645}

The design principle we have presented relies primarily on two ingredients: delocalization of excitons and near-resonant modes. These conditions can be found in several biological LHCs. For illustration, consider the light-harvesting complex Phycocyanin-645 (PC645), found in the cryptophyte algae \textit{Chroomonas} CCMP270. Cryptophytes are noteworthy for their ability to absorb solar energy even under low-light conditions. The PC645 system contains a pair of energetically similar and strongly coupled sites (called DBV C and DBV D) \cite{mirkovic07a,collini10a,pachon11a}, leading to a delocalized exciton dimer. As well, vibrational structures are believed to be important for understanding experimentally observed coherences \cite{turner12a,richards13a}, giving the second ingredient of the design principle.

\begin{figure}[!t]
\centering
\includegraphics[width=\columnwidth]{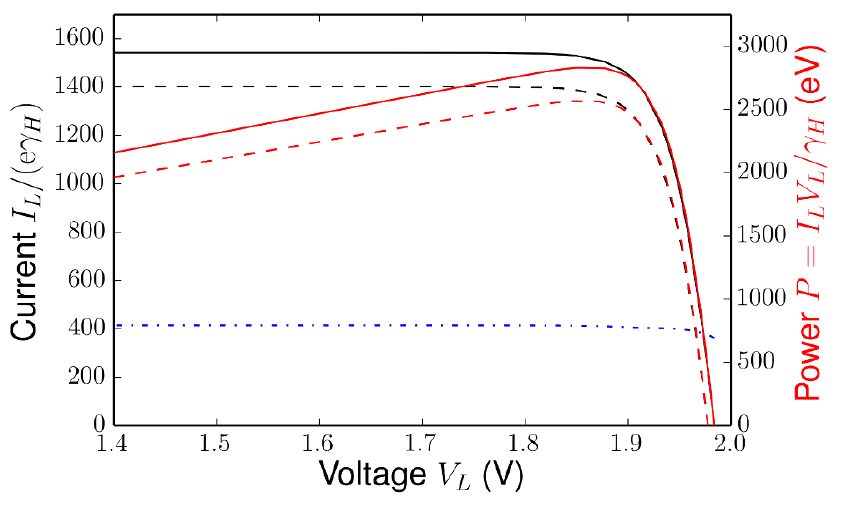}
\caption{Numerical I-V characteristic (upper black curve pair) and power (middle red curve pair) of the PC645-inspired system with (solid lines) and without (dashed lines) coherent exciton-mode interactions. The lower blue dash-dotted curve represents the magnitude of $I_\textrm{coh}$ up to n=5. The maximum power enhancement is $\sim 10.6\%$ when vibrational coupling is included.}
\label{fig:PC645IV}
\end{figure}

We consider a QHE model similar to above, but within a parameter range connected to the PC645 system. This model has four levels: a ground state, a coupled dimer pair, and an extra uncoupled lowest excited state based on the chromophore site PCB 158 D, which represents one of the major transfer pathways \cite{marin11a} out of the DBV C/D dimer (the other pathways have similar energies). The energy and coupling values are taken from \cite{mirkovic07a,pachon11a}. We also include quantized vibrational modes at sites 1 and 2, each with a frequency, based on the analysis of \cite{richards13a}, which is near-resonant with the excitonic splitting. A full list of parameter values can be found in Table \ref{tab:modelparameters}. The numerical I-V and power curves are presented in Fig. \ref{fig:PC645IV}, showing a maximum power enhancement of $\sim 10.6\%$.

\subsection{Comparison to artificial systems} 

There are a number of theoretical models in the literature for bio-mimetic or bio-inspired light-harvesting systems. We concentrate on proposals which use a heat engine approach, and which focus on the potential influence of quantum effects. In \cite{dorfman13a}, the authors propose a QHE model for the photosynthetic reaction center. In this system, the two excited states are (near-)degenerate in energy and interact with one another via noise-induced coherent effects. They assume the excited states share a common environment, but do not include any quantized modes. Their final level scheme is fairly similar to ours, though it is completely excitonic, while ours is combined excitonic-vibrational. Using a variational calculation, they find that the noise-induced coherence can increase the efficiency of charge separation in the reaction center or photocurrent in an artificial solar cell. They also use I-V characteristics to quantify the performance of the system at steady state. Unfortunately, the variational approach seems to allow non-physical evolution \cite{chin13b}. Thus, their proposed model, while conceptually interesting, rests on unsteady physical ground. This is one reason we have adopted a Lindblad master equation in this work. Though it might lack an underlying microscopic description, we avoid any strictly unphysical evolutions.

In \cite{chin13b}, the authors propose their own biologically inspired quantum heat engine photocell system. Although their basic level scheme resembles that of \cite{dorfman13a}, containing two energetically degenerate excited states, they rely on conceptually simpler mechanisms for enhancement. Quantum interference of dipole moments leads to new delocalized exciton states. These delocalized excitons have more favourable interactions with the underlying thermal reservoirs, leading to enhancements of up to 30\% in the steady state I-V and power curves.  

The starting point for our prototype model (before vibrations) is very similar to this photocell, including a bright and dark state which are, respectively, strongly and weakly coupled to the incoming energy. Of course, once vibrations are included, our prototype becomes more involved. From our model, we draw complementary conclusions to \cite{chin13b} about the strong functional importance of delocalization. The delocalization angle $\theta_J$ modulates the Jaynes-Cummings-type interaction sinusoidally (Eq. (\ref{eq:part_deloc_Ham})), and partially delocalized excitons had the most striking negative effects on our system's performance. Thus, it seems that the formation of strongly delocalized exciton states can be a crucial ingredient for enhancing light-harvester performance.


\section{Conclusion}

We have outlined the quantum advantage offered by coherent vibronic evolution in light-harvesting systems. Jaynes-Cummings interactions between the electronic and vibrational subsystems open up alternate, coherence-mediated, pathways for excitation transfer, allowing the total system to deliver energy faster than possible by incoherent thermal processes alone. Our results build a quantitative link between coherent vibronic evolution and functional advantage. The basic design principle was illustrated for an idealized prototype, for several non-ideal situations, and tested with parameters motivated by a biological LHC. Of course, the connection between experimentally observed quantum coherence and improved functionality in biological LHCs remains strongly debated. The LHC example above demonstrates, with experimentally relevant values, that vibronic coherence can indeed enhance performance within parameter regimes relevant for biological light-harvesters. The conclusive verification of these mechanisms in biological systems requires further theoretical and experimental investigation. As well, the presented design principle should not be expected to be universal. Depending on the system and its environmental constraits, the vibronic mechanism provides but one tool of many that can be used to achieve the goal of harvesting as much energy as possible. In any case, the design principle presented here can also serve to inspire vibronic-based designs for artificial light-harvesting systems.

\begin{acknowledgments}
Numerical computations were performed using the python package QuTiP \cite{johansson12a}.
We thank Felipe Caycedo-Soler for helpful discussions and acknowledge support from the ERC Synergy grant BioQ, the EU STREP PAPETS and QUCHIP, the DFG SFB TR/21 and an Alexander von Humboldt professorship.

SH thanks Alexandra Olaya-Castro and Edward O'Reilly for their hospitality in London, where the preliminary version of this work was presented, and for sharing unpublished results on the thermodynamical analysis of the effect of quantized vibrations, as described in a chapter of E. O'Reilly's PhD thesis \cite{oreillyThesis}.
\end{acknowledgments}


\appendix*

\section{Quantum advantage for a model with optimum incoherent transport}

The design principle outlined in this paper relies on the interplay of three primary components. These are incoherent baths, delocalized electronic states, and coherent electron-mode evolution via a Jaynes-Cummings interaction. The incoherent baths provide directionality through driving and damping; the delocalized excitons allow for transfer to even take place at all given the local form of the baths; and the Jaynes-Cummings interaction increases the overall rate of transfer by offering alternate pathways. Clearly, the design principle will work best when neither the incoherent nor the coherent processes dominate. However, in our idealized model, the possibility exists to turn up the coupling $\gamma_C$ to the cold bath until incoherent transport speeds dominate coherent oscillation frequencies. In such a regime, the coherent exciton-mode evolution would operate on too slow a timescale to provide any quantum advantage. Of course, one could also increase the coupling $g$ to the expectional vibrations in the model until the quantum nature of
evolution again provided an advantage. While our idealized model allows for this kind of parameter manipulation, doing so is not really in the spirit of what is happening physically.

We would expect that our prototype model only describes the physical system well for some appropriate range of parameters. In particular, we should not expect the physics to be the same for all bath strengths. When the bath strength dominates over the electronic degrees of freedom, the preferred electronic states which undergo transitions will be effectively relocalized \cite{fassioli14a}, since the underlying environmental interaction is typically thought to be itself local.
But the more localized the preferred electronic states, the less a localized interaction of the form of Eq. \ref{eq:HI_pt} can actually facilitate transitions between them. Thus, bath-induced relocalization can suppress the actual transition rates when the bath strength goes beyond some optimal value.

\begin{figure}[!t]
	\includegraphics[width=\columnwidth]{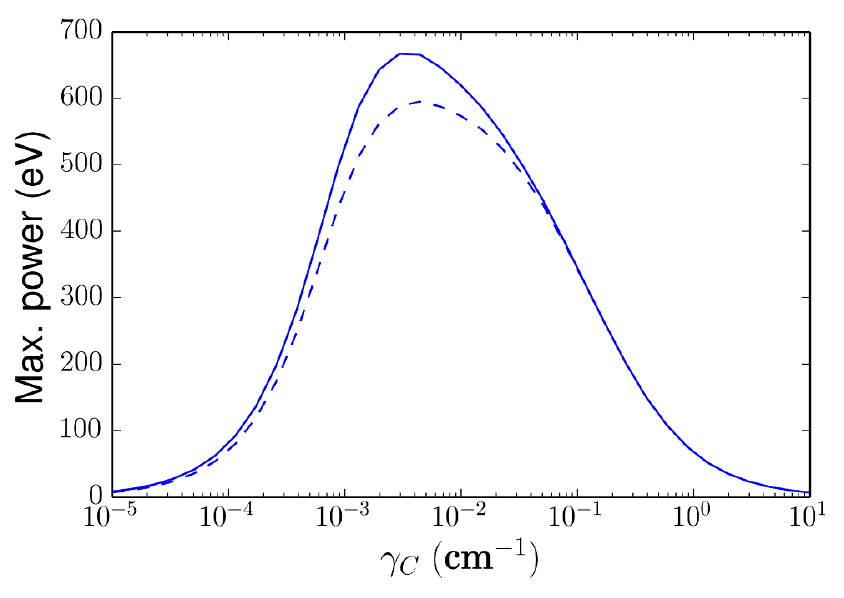}
	\caption{a) Maximum power point versus $\gamma_C$ in a model with bath-induced relocalization, both with (solid lines) and without (dashed lines) coupling to the quantized modes. In an appreciable region around the optimum, the coupled modes provide higher maximum power than is possible for any choice of bath strength parameter $\gamma_C$.}
	\label{fig:optcompare}
\end{figure}

We model this situation by introducing a background-induced localization angle $\theta_B:=\arctan(\frac{\alpha}{\gamma_C})$, where $\alpha=1~\mathrm{meV}$ is a constant (this should be distinguished from the delocalization angle caused purely by electronic coupling). The relocalized excitonic states are
\begin{align}
	\ket{\varepsilon_1(\theta_B)}:=
	\cos(\tfrac{\theta_B}{2})\ket{s_1}+\sin(\tfrac{\theta_B}{2})\ket{s_2},\\
	\ket{\varepsilon_2(\theta_B)}:=
	\sin(\tfrac{\theta_B}{2})\ket{s_1}-\cos(\tfrac{\theta_B}{2})\ket{s_2},
\end{align}
with $\ket{\varepsilon_3(\theta_B)}=\ket{s_3}$ and $\ket{\varepsilon_0(\theta_B)}=\ezero$ as before.
When $\gamma_C\ll \alpha$, the excitons are completely delocalized, while if $\gamma_C\gg \alpha$, they become localized. Finally the Lindblad transition operators are modified to take place between the $\theta_B$-exciton states with new rates $\gamma_C(\theta_B) = \gamma_C\sin^2(\theta_B)$ which are weighted by the degree of localization. As well, the bath occupations $\overline{n}_{Cij}(\theta_B)$ will also depend on $\theta_B$ since the energy differences for Eq. (\ref{eq:planckdist}) are modified by the delocalization angle. In summary, the Lindblad components are exactly as in the earlier table, except with the replacements $\ket{\varepsilon_i}\rightarrow\ket{\varepsilon_i(\theta)}$, $\gamma_C\rightarrow \gamma_C(\theta)$, and $\overline{n}_{Cij}\rightarrow\overline{n}_{Cij}(\theta)$.

Numerically, we determine the steady state of this incoherent transport model for various values of the coupling parameter $\gamma_C$, first with no quantized vibrations. We show the maximum power point for each value of $\gamma_C$ in Fig. \ref{fig:optcompare} (dashed curve). As expected, the incoherent model has an optimal choice of $\gamma_C$, with relocalization effects suppressing the power after this optimum. We then add in quantized modes at sites 1 and 2, at frequency $\hbar\omega_m = 200 \invcm$. The max power points for this situation are given by the solid curve in Fig. \ref{fig:optcompare}. It is clear that even if we are at the regime of optimal incoherent transport, the inclusion of coherently coupled vibrations provides an additional quantum advantage to the power of the light-harvesting system.

\bibliography{vibronic_qhe_refs}

\end{document}